\documentclass[a4paper]{article}
\usepackage{graphicx}
\usepackage{amsmath}
\begin{document}

\begin{center}
  
  \hbox{}\vspace{0.5cm}
  {\large \bf Active percolation analysis of pyramidal neurons of somatosensory cortex: A comparison of
  wildtype and p21H-ras$^{\rm Val12}$ transgenic mice}\\
  \vspace{0.5cm} {\sf Luciano da Fontoura Costa} and
  {\sf Marconi Soares Barbosa}\\
  Instituto de Física de São Carlos \\
  Universidade de São Paulo Caixa Postal 369 CEP 13560.970 São Carlos, SP, Brazil\\
  {\sl e-mail: marconi@if.sc.usp.br}\\
  \vspace{0.5cm} {\sf Andreas Schierwagen }\\
  Institute for Computer Science\\ University of Leipzig, D-04109 Leipzig, Germany\\
  \vspace{0.5cm} {\sf Al\'an Alp\'ar and Ulrich G\"artner and Thomas Arendt}\\
  Department of Neuroanatomy\\ Paul Flechsig Institut for Brain Research, University of Leipzig,
  D-04109 Leipzig, Germany\\

\end{center}

\vspace{1.0cm}

\begin{abstract}
  This article describes the investigation of morphological variations
  among two set of neuronal cells, namely a control group of wild type
  rat cells and a group of cells of a trangenic line.  Special
  attention is given to sigular points in the neuronal structure,
  namely the branching points and extremities of the dendritic
  processes. The characterization of the spatial distribution of such
  points is obtained by using a recently reported morphological
  technique based on forced percolation and window-size compensation,
  which is particularly suited to the analysis of scattered points
  presenting several coexisting densities. Different dispersions were
  identified in our statistical analysis, suggesting that the
  transgenic line of neurons is characterized by a more pronounced
  morphological variation. A classification scheme based on a
  canonical discriminant function was also considered in order to
  identify the morphological differences.
\end{abstract}

\section{Introduction} \label{sec:intro}

During brain development, neurons form complex dendritic and axonal arbors
that reach a characteristic pattern and size~\cite{Feldman:1984}. The
development of arbor shape is partly determined by genetic factors and partly
by interactions with the surrounding tissue
(e.g.~\cite{Miller:2003,Scott:2001,Wong:2002}). Dendritic growth is known to
be responsive to various environmental signals, including synaptic activity
and guidance molecules~\cite{McAllister:2000,Scott:2001,Whitford:2002}. The
impact of neurotrophins upon cortical neurons has gained special interest,
since they are essential for neuronal development, as well as for the
maintenance of functional stability and plasticity in the adult nervous
system~\cite{Thoenen:1995,Davies:2000,Huang:2001,McAllister:2002}.

Both neuronal activity and neurotrophins guide dendritic development via
GTPase-dependent mechanisms~\cite{McAllister:2000,Li:2002,Sin:2002}.  Recent
studies have drawn attention onto the small G-protein p21Ras in dendritic
growth~\cite{Holzer:2001a,Holzer:2001b,Huang:2003}. Like other GTPases, p21Ras
regulates the phosphorylation of downstream kinases which trigger cascade
mechanisms resulting in the regulation of enzymatic activities, ionic
channels, cellular morphology and gene
expression~\cite{Ahn:1993,Nishida:1993,Heumann:1994}.

Effects of p21Ras have been shown to be essential for the normal functioning
and plasticity of both the developing~\cite{Noda:1985,
  Bar-Sagi:1985,Guerrero:1986} and the adult nervous
systems~\cite{Brambilla:1997,Silva:1997,Moore:2000}.  Furthermore, enhanced
expression of p21Ras is associated with neuronal restructuring after lesion or
in the context of neurodegenerative
diseases~\cite{Phillips:1994,Gartner:1995,Gartner:1999}.

The effect of constitutively active p21Ras protein upon the mature nervous
system has been previously investigated in p21H-Ras$^{\rm Val12}$ transgenic
mice~\cite{Heumann:2000,Holzer:2001a,Holzer:2001b}. In this model the
expression of transgenic p21H-Ras$^{\rm Val12}$ starts postnatally around day
15, when neurons are postmitotic and the majority of synaptic contacts has
been established. In par ticular cortical pyramidal neurons express the
transgenic construct at high levels. The volume of the cerebral cortex of
these transgenic mice is increased by approximately 20\% compared to wildtype
mice.  This increase has been attributed to the enlarged volume of the
cortical pyramid al cells, but not to a raise in the number of
neurons~\cite{Heumann:1996,Heumann:2000,Holzer:2001a,Holzer:2001b}.

Dendritic morphology of pyramidal cells is very heterogeneous, not only in
different areas (e.g.~\cite{Elson:2002}) but also within the same region
(e.g.~\cite{Duan:2002}). To set the population of neurons for investigation,
only commissural neurons of layers II/III of the primary somatosensory cortex
have been analysed in this study. By means of this restriction the analysis of
a relatively homogeneous subpopulation of pyramidal neurons became possible.

\begin{figure}[hb]
\begin{center}
\includegraphics[scale=0.3,angle=0]{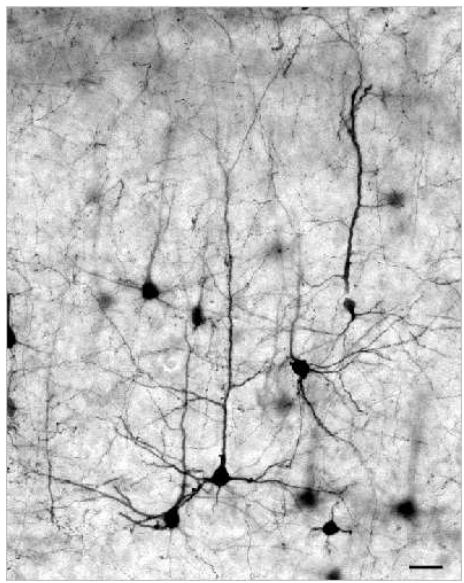}

\caption{Retrogradely pyramidal neurons in layers II/III of the
  primary somatosensory cortex. Scale bar: 50 $\mu m$.~\label{fig:pat} }   
\end{center}
\end{figure}

In a previous study we found that enhanced p21Ras activity results in a
dramatically enlarged dendritic tree.  In both cortical layers II/III and V,
the total surface area and the total volume of dendritic trees is greatly
increased. This is mainly caused by increased dendritic diameter and by the
appearance of additional segments~\cite{Alpar:2003}.

On the other hand, topological complexity of pyramidal neurons in layers
II/III appeared hardly affected: Sholl analysis of both basal and apical
dendrites revealed no differences between transgenic and wildtype mice
regarding any parameters considered, i.e. numbers of intersections, branching
points and tips~\cite{Alpar:2003}.

These results suggest a rather proportional increase of dendritic tree size,
without distinct changes in the space-filling properties. The aim of the
present study was to substantiate these findings.

In the present work we consider the use of a new procedure, known as
percolation transform~\cite{Costa1,Costa2}, which is particularly useful for
characterization of spatial density of distributed points. This framework
represents an enhanced alternative to the multiscale fractal analysis reported
before~\cite{Costa3}.  The percolation transform procedure consists of two
steps. The first is to induce an active percolation dynamics by performing an
exact dilation of the points.  During this process, the number of mergings
between points and connected groups of points is recorded for each dilation
radius. Having recorded the accumulative number of mergings as a function of
scale, an essential normalization procedure is performed in order to
compensate for sparser distributions which would have produced a much weaker
signal otherwise.

The overall enhanced sensitivity is a direct consequence of recording only
mergings or shocks between the growing clusters, an abrupt phenomena which may
occur at any dilation radius, rather than the smooth increase in area as
considered in the (multiscale) fractal analysis \cite{Costa3}. This enhanced
type of analysis is particularly suitable in the present case because the
percolation transform approach is independent of size-related parameters like
area and volume of the neuronal cells.  We argue that changes in the global
character of the percolation transform curves, derived from the reference
points (i.e. dendrite extremities and branch points) of the dendrites of
pyramidal neurons due to transgenic activation of p21Ras in the primary
somatosensory cortex of mice, correlate with changes in the complexity of
neuronal morphology.

\section{Materials and methods}

\subsection{Biological}

We used data derived from experiments with three transgenic mice aged
nine months (see~\cite{Heumann:1996,Heumann:2000} for the
establishment of the transgenic mouse line), as well as with three
wildtype mice of the same age. All experimental procedures on animals
were carried out in accordance with the European Council Directive of
24 November 1986 (86/609/EEC) and had been approved by the local
authorities. All efforts were made to minimise the number of animals
used and their suffering.

Mice were anaesthetised with Hypnomidate$^{\text{\textregistered}}$(Janssen-Cilag)
(1 ml/40 g body weight intraperitoneally) and positioned in a stereotaxic
apparatus. Prior to incision 0.1 ml Xylonest$^{\text{\textregistered}}$ (Astra) was
injected below the skin, the left parietal bone was partially removed, Using a
Hamilton syringe for unilateral injection, Biotinylated Dextran Amine (1 $µ$l
20\% BDA, 10.000 nominal molecular weight(MW), Molecular Probes) was mechanically delivered into the
corpus callosum 1mm lateral to the midline and 0.5 mm caudal to the bregma,
1mm deep from pial surface according to the stereotaxic atlas of Frankin and
Paxinos (1997). Seven days after surgery the animals were intracardially
perfused in deep anaesthesia, first with saline (0.9 \% NaCl) for 1-2 min and
then with a fixative containing 4\% paraformaldehyde in 0.1 M phosphate buffer
(pH 7.4) for 30 min. The brains were postfixed in the same solution overnight,
immersed in 30\% sucrose for another 24 hours, and sectioned (160 $µ$m) in
the coronal plane on a cryostat. The free-floating sections were extensively
washed in 0.05 M Tris Buffered Saline (TBS, pH 7.4), and then reacted with the
avidin-biotin-peroxidase complex (1:100 in TBS, VECTASTAIN$^{\text{\textregistered}}$
Elite ABC kit, Vector Laboratories, Inc.) at room temperature for two hours.
BDA was visualised by using 3,3´-diamino-benzidine (Sigma, 0.025 \%)
intensified with nickel-ammonium sulphate (Merck, 0.05 \%) in the presence of
0.001 \% hydrogen peroxide, diluted in TBS. Sections were mounted onto
gelatine coated glass slides, dehydrated and covered with DPX (Fluka, Neu-Ulm,
Germany).

The retrogradely labelled pyramidal cells were reconstructed using
Neurolucida$^{\text{\textregistered}}$ (MicroBrightField, Inc.), see
Figure~\ref{fig:pat}. The system allowed accurate tracing of the cell
processes in all three dimensions and continuous adjustment of the dendritic
diameter with a circular cursor. A motorized stage with position encoders
enabled the navigation through the section in the $xyz$ axes and the accurate
acquisition of the spatial coordinates of the measured structure. All visible
dendrites were traced without marking eventual truncation of smaller dendritic
sections. This may have led to certain underestimation of the dendritic tree,
especially in transgenic mice with a larger dendritic arbour.  To gain both
optimal transparency for optimal tracing facilites and at the same time a
possibly complete neuronal reconstruction, sections of 160 $µ$m thickness were
used. Thicker sections allowed only ambigous tracing of thinner dendritic
branches. Shrinkage correction (300 \%) was carried out in the z axis, but not
in the xy plane, because shrinkage was negligible in these dimensions
(~10 \%).

\begin{figure}[hbt]
\begin{center}
\includegraphics[scale=0.2,angle=0]{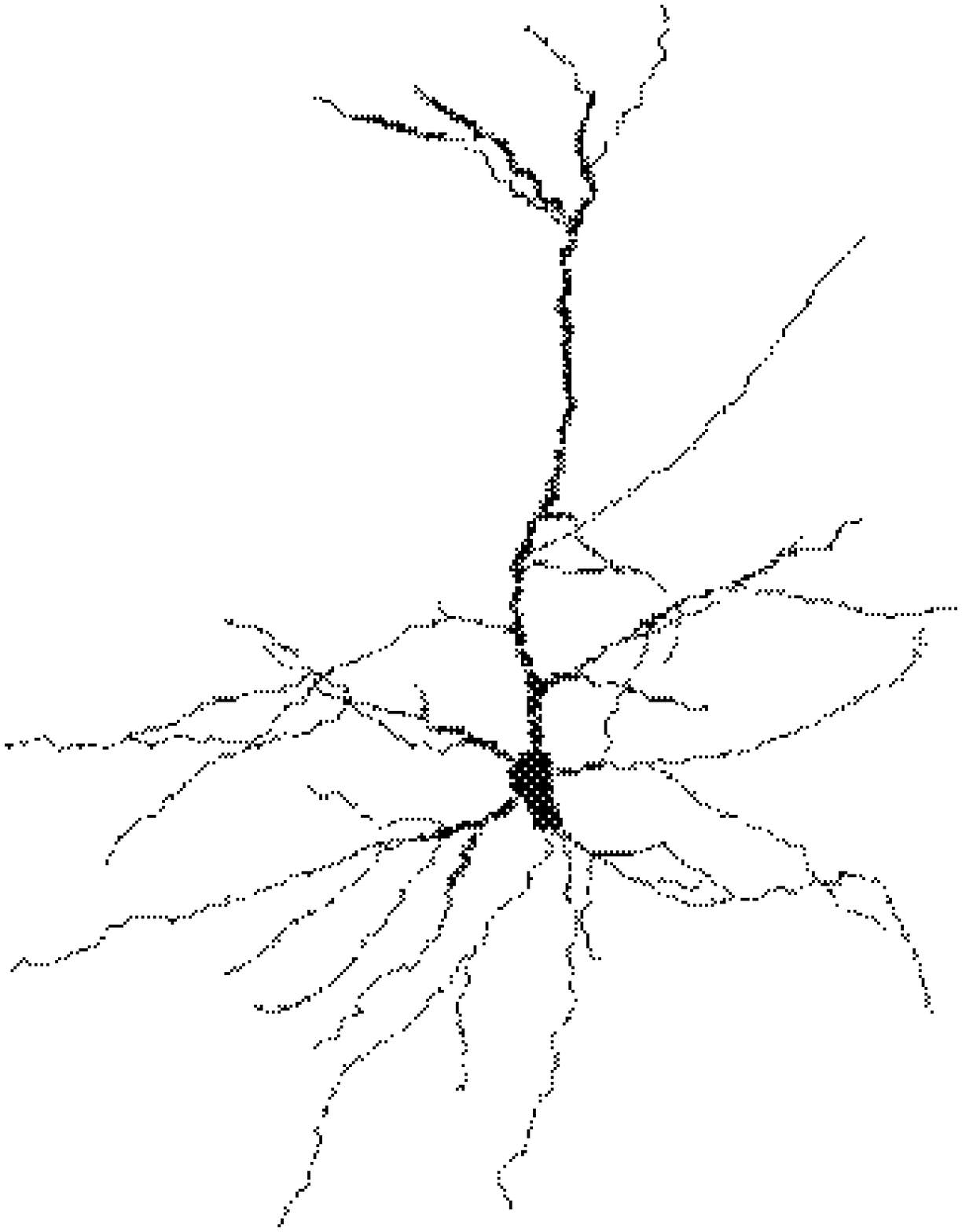}\hspace{1cm}\includegraphics[scale=0.2,angle=0]{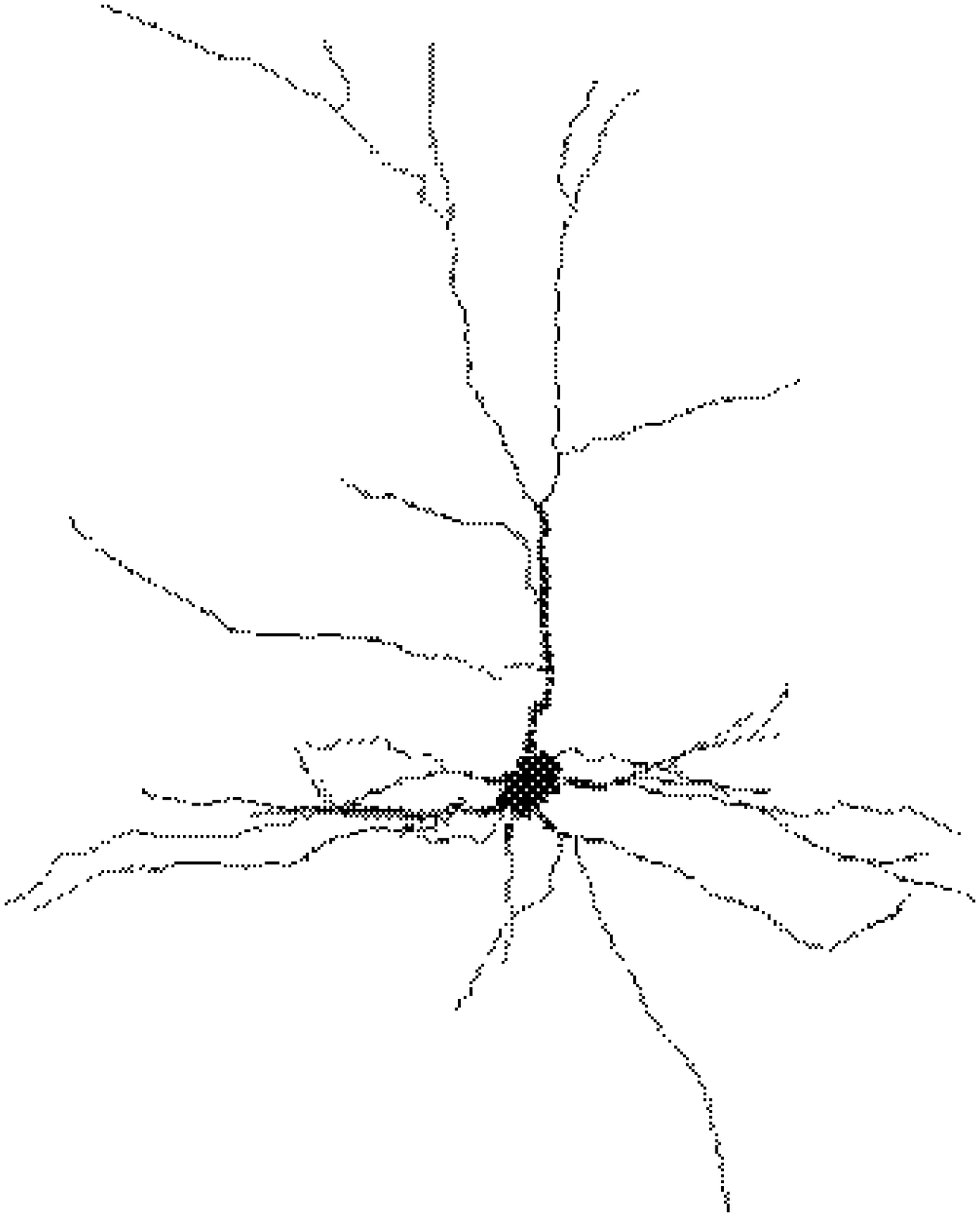}
\caption{Pyramidal cells rendered with Cvapp. Displayed are random examples of each transgenic neuron (cell SE8, left) and
  wildtype neuron (cell WT19, right).\label{fig:swc} }
\end{center}
\end{figure}

The morphology files created by Neurolucida$^{\text{\textregistered}}$ need to be
edited and converted to a simpler file format before they can be used for
further analyses. We used for this purpose the freely available program
Cvapp~\cite{Cannon:2000}. Cvapp is a cell viewing, editing and format
converting program for morphology files, see Figure~\ref{fig:swc}. It can be
also used to prepare structures digitized with Neurolucida$^{\text{\textregistered}}$
software for modeling with simulators like Neuron and Genesis. There are
helpful guidelines on how to use cvapp to convert Neurolucida ASCII files to
the SWC format describing the structure of a neuron in the simplest possible
way~\cite{Jaeger:2000}.

Each line encodes the properties of a single neuronal compartment. The format
of a line in a SWC file is as follows: n T x y z R P.  In turn, these numbers
mean:  (1) an integer label (normally increasing by one from one line to the
next) that identifies the compartment, (2) an integer that represents the type
of neuronal segment (0-undefined, 1-soma, 2-axon, 3-dendrite, 4-apical
dendrite, etc.)  , (3)-(5) xyz coordinate of compartment , (6) radius of
compartment, (7) parent compartment (defined as -1 for the initial
compartment). The sample of neurons edited, converted and analyzed (see below)
comprised 28 of wildtype neurons and 26 of transgenic neurons.

\subsection{Computational}

The concept of the percolation transform~\cite{Costa1,Costa2} is described in
the following. As mentioned in Section~\ref{sec:intro} this method uses of a
mechanism known as Minkowski dilation. Given a set of points $S$ in $R^3$, its
Minkowski dilation by a radius $r$ corresponds to the union of closed balls of
radius $r_i$ placed over each point in $S$, and the volume of the dilated set
is represented as $V(r_i)$. In order to ensure enhanced precision and
computational efficiency, only those radiuses corresponding to the viable
distances in the orthogonal lattice, the so-called exact
distances~\cite{Costa:1999}, are taken into account.

\begin{figure}[ht]
\begin{center}
\includegraphics[scale=0.7,angle=0]{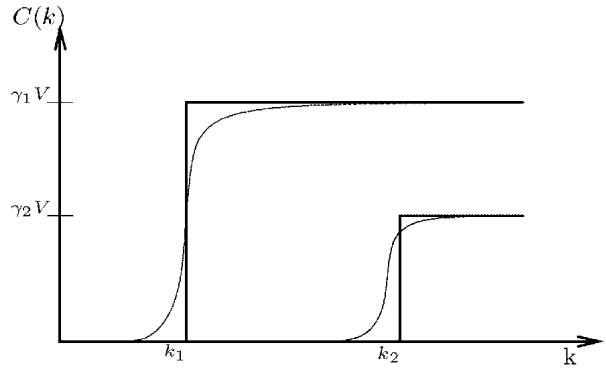}
\caption{The smooth curve shown corresponds (schematically) to the
  accumulative mergings as a function of the characteristic scale for a
  Poisson distribution of points. The unit-step functions shown come from
  idealized assumptions.~\label{fig:shock}}
\end{center}
\end{figure}

Consider now that the set of points $S$ is generated from a Poisson 
distribution with density $\gamma$ inside a cubic window of side
$L$. Let $C(d_i)$ be a function that counts the number of mergings
between growing clusters at a specific distance $d_i=2r_i$ as the
whole set of points undergoes a dilation by radius $r_i$. In order to allow
scale uniformity we make the change of variables $k_i=\ln(d_i)$. A graphic
representation of this function is shown in Figure~\ref{fig:shock} for two
different values of density $\gamma_1$ and $\gamma_2$. One can see that there
is a characteristic scale for each density, where most of the merging dynamics
takes place before reaching a plateau. The total
number of points in $S$ is given by $\gamma_1 V$
or $\gamma_2 V$ as illustrated by Figure~\ref{fig:shock}. The idealized
scenario where this dynamics occurs is represented by the unit step
functions. In  our model, the position of such a transition is taken as the
characteristic  scale for the set $S$.

%% Consider now that the set of points $S$ is generated from a Poisson
%% distribution with a density $\gamma$ inside a cubic window of side
%% $L$. Let $C(d_i)$ be a function that counts the number of mergings
%% between growing clusters at a specific distance $d_i=2r_i$ as the
%% whole set of points undergoes a dilation by radius $r_i$. A graphic
%% representation of this function is shown in Figure~\ref{fig:shock} for
%% two diferent values of density $\gamma_1$ and $\gamma_2$. One can see
%% that there is a characterisc scale where most of the merging dynamics
%% takes place before reaching a plateau with height corresponding to the
%% total number of points $\gamma V$.  The idealized scenario where this
%% dynamics occurs immediately along the density axis ($k_1$ or $k_2$ in
%% Figure~\ref{fig:shock}) is represented by the unit step functions. In
%% our model, the position of such a delta is taken as the characteristic
%% scale for the set $S$.

Next, we need to relate the caracteristic scales to the densities of
the generated Poisson set of points. We argue that the distance where
all points suddenly colapse into a unique cluster is in fact the mean
nearest neighbor distance. Knowing the nearest neighbor distance
distribution~\cite{torquato:1990} for the three-dimensional space, the
mean nearest neighbor distance as a function of the density of the
cloud can be expressed as

\begin{equation}\label{eq:near}
\bar{d}=\frac{0.554}{\sqrt[3]{\gamma}}.
\end{equation}

\begin{figure}[ht]
\begin{center}
\includegraphics[scale=0.5,angle=0]{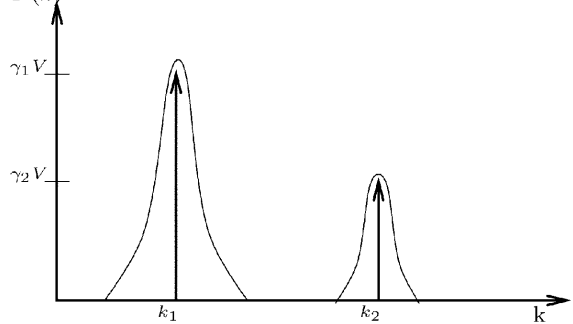}
\caption{The derivative of the cumulative merging
  function, as a function of the characteristic scale. Schematically as before
  the smooth curves stands for Poisson clouds and the Kronecker functions,
  represented as upward arrows, are associated with the idealized
  model.~\label{fig:gauss}}
\end{center}
\end{figure}

The derivative of the cumulative number of mergings function, $C(d_i)$, is
show schematically in Figure~\ref{fig:gauss}. The smooth curve represents the
behaviour of the shock dinamics for a Poisson model, while the unormalized
Dirac deltas stands for the idealized model. Note that the rate of change of
the cumulative number of mergings for higher scales tends to decrease fastly.
As we intend to use these characteristic scales to detect multiple densities
coexisting in the same set of points, we need a normalization procedure in
order to enhance the signal for sparser distributions of points, therefore
assigning the same importance to all scales. To implement such a normalization
we take as reference the maximum value of the rate of change for one specific
reference scale $k_j$. The function we need should always give the same value
irrespectively of the characteristic scale we are dealing with, as in the
following equation

\begin{equation}\label{eq:normal}
Q(\bar{k}_i)=\frac{P(\bar{k}_i)}{\Delta\bar{k}_j}e^{(3\bar{k}_i-3\bar{k}_j)}.
\end{equation}

This normalization procedure is the last step in the percolation
approach. In the case of an experimental cumulative function obtained
from a Poisson simulation with specific density,
Equation~\ref{eq:normal} will produce the same intensity for any
density value. When dealing with an experimental cumulative function
involving an unknow mixture of densities, Equation~\ref{eq:normal}
will yield detect the multiple characteristic distances actually
occuring in the sampled space.

\begin{figure}[ht]
\begin{center}
\includegraphics[scale=0.5,angle=-90]{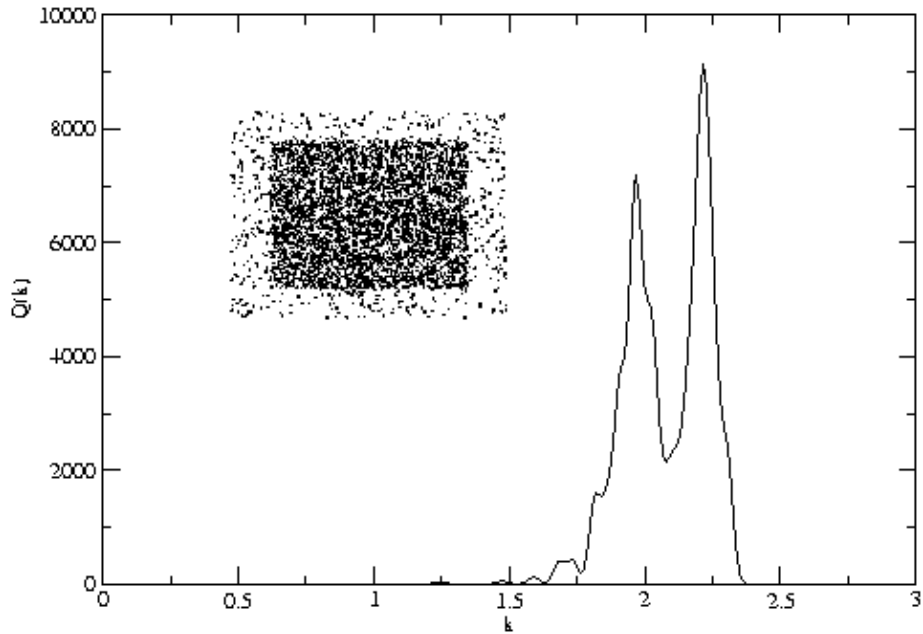}
\caption{An example showing the resolving ability of the proposed
procedure for the pattern show at the inset. The two coexisting
densities are clearly detected by the peaks occuring at their
characteristic scales.~\label{fig:cubes}}
\end{center}
\end{figure}

We illustrate the potential of this procedure regarding coexisting densities
of spatially distributed points for the following experimental example.
Figure~\ref{fig:cubes} presents an example pattern (inset) and the
corresponding percolation transform curve. The example pattern consists of two
Poisson clouds with distinct densities. The percolation curve for this example
shows clearly the two peaks corresponding to the characteristics scales
associated with the two different densities. Note that the difference in
height of those peaks reflects the relative area occupied by each clowd.

In the present work we consider the spatial distribution of the singular
points of the dendritic trees, i.e. dendritic branch points and extremities.
Such an approach (see~\cite{Barbosa:2003b,Sholl:1958}) has shown to be an
effective route when dealing with subtle geometrical changes while considering
a single morphological/physiological class of neuronal cells.

\section{Results}

\begin{figure}[hb]
\begin{center}
\includegraphics[scale=0.4,angle=-90]{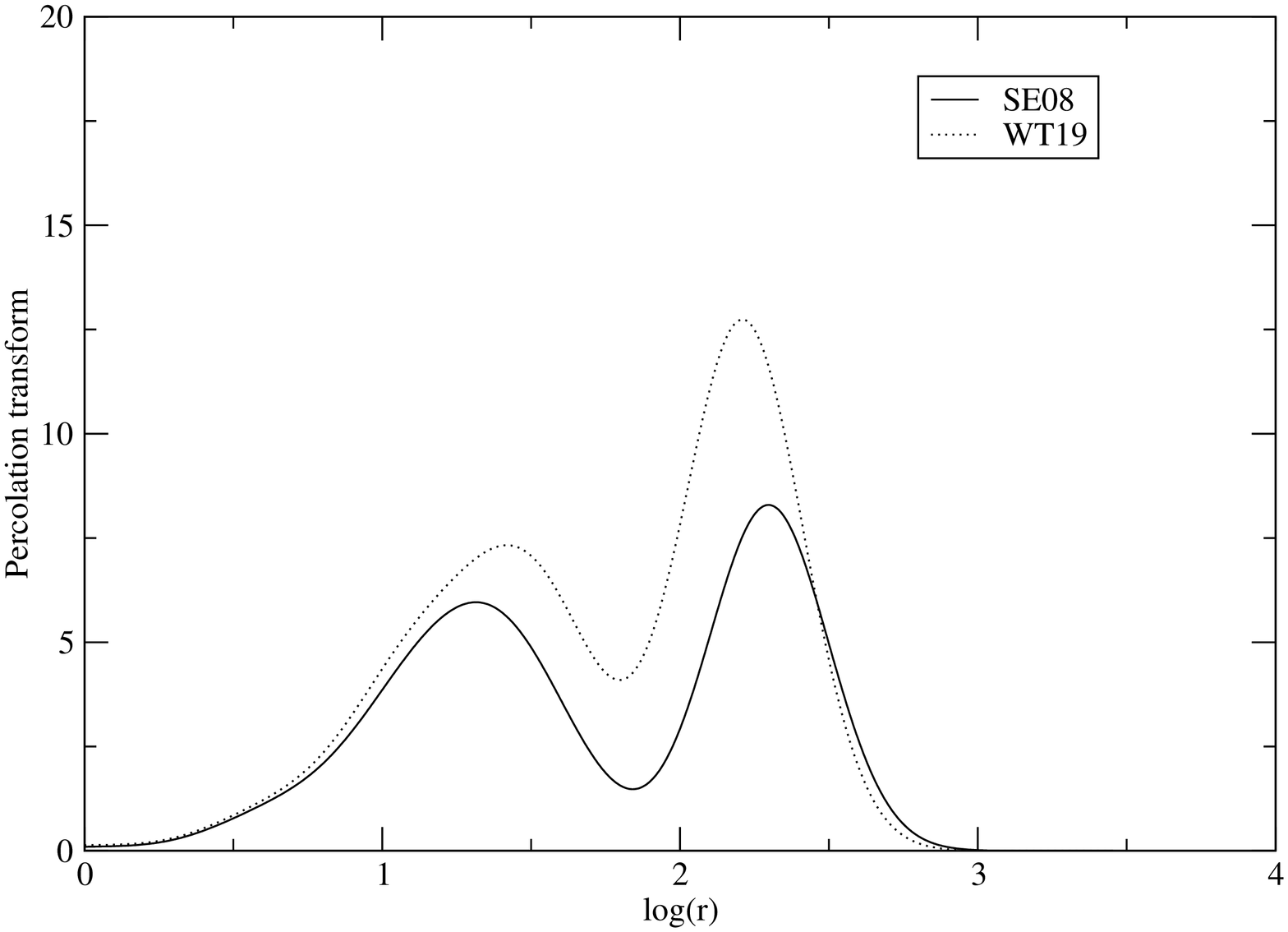}
\caption{Example of the percolation curve for the two types of
  neuronal cells presented in Figure~\ref{fig:swc}.~\label{fig:example}}
\end{center}
\end{figure}

\begin{figure}[hb]
\begin{center}
\includegraphics[scale=0.4,angle=-90]{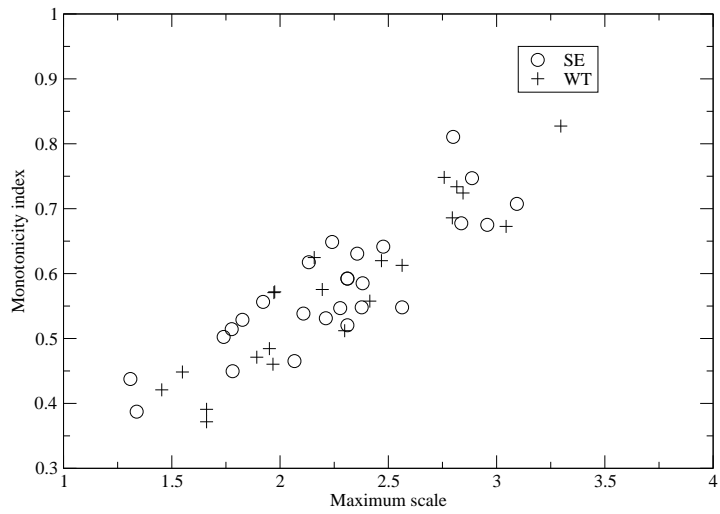}
\caption{A scatter plot defined by the monotonicity index (vertical
axis) and the scale (horizontal axis) at which the maximum of the
percolation transform curve ocurs.~\label{fig:scatter} }
\end{center}
\end{figure}

Figure~\ref{fig:example} shows the resulting percolation transform
curve for the two selected cells of Figure~\ref{fig:swc}. The overall
profile of these curves were the same, showing two proeminent peaks at
clearly distinguishable scales~\footnote{Note that the scale axis is
the logarithm of the actual distance among the points.}. Although we
can see that these two single cells presents a clearly distinc global
features in their signature curve, this property is masked by the
statistical variabitities of the whole set of cells, as shown by the
scatter plot in Figure~\ref{fig:scatter}. This scatter plot is an
example of various quantile-quantile graphs we produced using global
properties of all percolation curves, such as those exemplified in
Figure~\ref{fig:example}. For this purpose, we select a few global
properties, namely the mean, the variance, the maximum value, the
scale at which the maximum value occur and the monotonicity
index\cite{Barbosa:2003a}. The scatterplot of Figure~\ref{fig:scatter} is
defined by the monotonicity index and the maximun scale, as an example
of the overall variability of shapes. Taken pairwise none of these
global measures leads to clear separation of the two types of cell
involved in this study.

\begin{figure}[hb]
\begin{center}
\includegraphics[scale=0.5,angle=0]{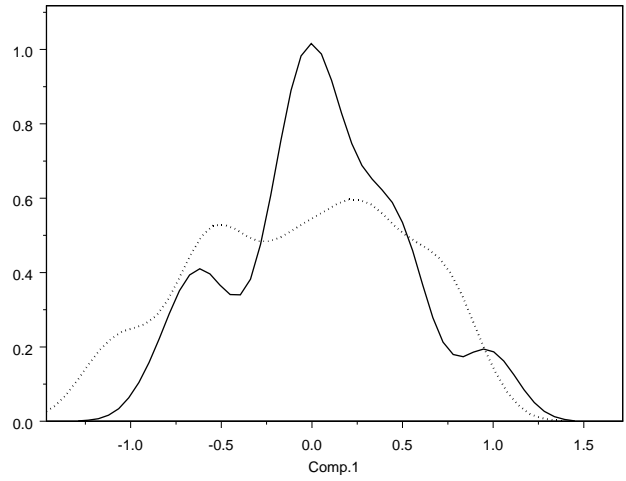}
\caption{The density profile of the principal component score for each
class in this experiment. One can see a very similar mean and
broader dispersion in the WT cells shown by the dotted line.~\label{fig:density}
}

\end{center}
\end{figure}

\begin{figure}[hb]
\begin{center}
\includegraphics[scale=0.5,angle=0]{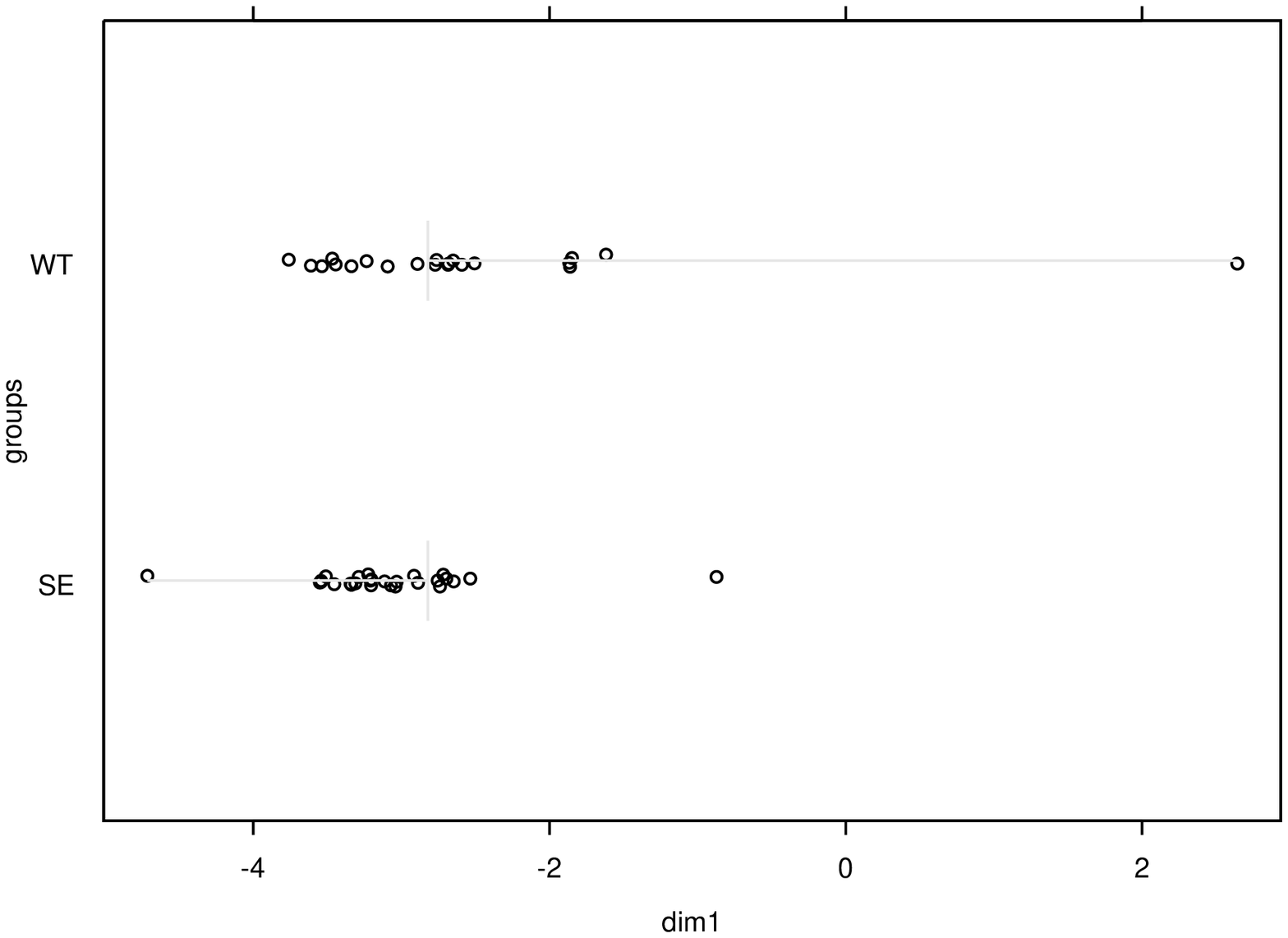}
\caption{The graphical result of canonical analysis, showing the
amount of discrimimation possible for both classes of neurons with the
one dimensional variable.~\label{fig:canographic} }
\end{center}
\end{figure}

Before ruling out the possible existence of a geometrical
characteristic that may be correlated with the genetic treatment, we
performed a more thorough statitical analysis of the data produced by
the percolation approach.  These consist of a principal component
analysis and a cononical discriminant analysis
\cite{McLachlan:1992}. Figure~\ref{fig:density} shows the density profile for the
principal component variable, showing a similar mean but a pronouced
diference in dispersion of the considered geometrical features of the
genetically treated cells. Table~\ref{tab:cross} shows the performance
of the those selected global features in providing a discriminant
function for the two type of cells. While there is visible potential
for classifyng one type of cells, namely the SE group, the
discriminant function has a marginal performance concernig the other
group, misclassifying almost half of them. This can be graphically
visualized in Figure~\ref{fig:canographic}.

\begin{table}[ht]
\begin{center}
\begin{tabular}{l|l|c|c|c}\hline\hline
    &  SE &  WT &    Error & Posterior.Error \\\hline
SE & 20 &  7 & 0.2592593 &     0.3891051   \\\hline
WT  &  9 & 12 & 0.4285714 &     0.4767072   \\\hline
Overall  &    &    & 0.6878307 &     0.4274310   \\\hline\hline
\end{tabular}
\end{center}
\caption{The result of classical discriminant analysis for the
measures considered in the scatterplot of
Figure~\ref{fig:scatter}. From this cross-validation table we observe
a somewhat better classifacation rate for type SE while for WT the
estimation is indecisive.~\label{tab:cross} }
\end{table}

\section{Discussion}

The present study is one of the few which uses 3D data on neuronal morphology
to quantitatively characterize the complexity and scaling properties of
different neuron types.  The morgphology of pyramical cells was investigated
in two sets of neurons, i.e. wildtype and p21H-rasVal12 transgenic mice. The
results obtained using principal component analysis indicates that the
transgenic neurons have a greater dispersion, as revealed by the density
profiles shown in~Figure~\ref{fig:density}. This finding may suggest that the
enhanced p21Ras activity in transgenic mice may lead to greater variety of the
cell morphological phenotype.

These results must be valuated with care, however. We know from other studies
(e.g . Fernandez et al. 1994) that a measure like fractal dimension for
example is not sufficient to differentiate between cell classes. Alone it does
not completely specify a cell´s morphology but increases classification
accuracy as an additional parameter for morphological classification among
several other parameters typically used for placing nerve cells in different
classes such as soma diameter, maximum dendritic diameter, number of branches
etc. The same situation may have occured here with the percolation transform
approach.  Although our findings suggest that transgenic mice pyramidal
neurons exhibt a more variable dendritic morphology than the corresponding
wild type neurons, this problem must be investigated in more detail. Clearly
such study will include a more extended statistical analyses of the neuronal
morphometrical parameters.

\section{Acknowledgments}
This study was partly supported by the Deutsche Forschungsgemeinschaft
(grant GA 716/1-1), FAPESP and CNPq.

\clearpage
\bibliographystyle{unsrt}
\bibliography{perctrans}

\end{document}